\documentclass[twocolumn,prl,preprintnumbers,amsmath,amssymb,superscriptaddress,aps]{revtex4}

\pdfoutput=1
\usepackage[font=footnotesize,labelfont=sf,bf,flushleft]{caption}
\usepackage{graphicx}
\usepackage[usenames,dvipsnames]{xcolor}
\usepackage{amsmath,bbm}

\begin{document}
\raggedbottom
 
\title{Full Rotational Control of Levitated Silicon Nanorods}

\author{Stefan Kuhn}
\email{stefan.kuhn@univie.ac.at}
\affiliation{University of Vienna, Faculty of Physics, VCQ, Boltzmangasse 5, 1090 Vienna, Austria}%Lines break automatically or can be forced with 
\author{Alon Kosloff}
\thanks{Co-first Author}
\affiliation{School of Chemistry, Tel-Aviv University, Ramat-Aviv 69978, Israel}
\author{Benjamin A. Stickler}
\affiliation{University of Duisburg-Essen, Lotharstra{\ss}e 1, 47048 Duisburg, Germany}
\author{Fernando Patolsky}
\affiliation{School of Chemistry, Tel-Aviv University, Ramat-Aviv 69978, Israel}
\author{Klaus Hornberger}
\affiliation{University of Duisburg-Essen, Lotharstra{\ss}e 1, 47048 Duisburg, Germany}
\author{Markus Arndt}
 \affiliation{University of Vienna, Faculty of Physics, VCQ, Boltzmangasse 5, 1090 Vienna, Austria}
\author{James Millen}
 \affiliation{University of Vienna, Faculty of Physics, VCQ, Boltzmangasse 5, 1090 Vienna, Austria}

\begin{abstract}
Optically levitated nano-objects in vacuum are amongst the highest-quality mechanical oscillators, and thus of great interest for force sensing, cavity quantum optomechanics, and nanothermodynamic studies. These precision applications require exquisite control. Here, we present full control over the rotational and translational dynamics of an optically levitated silicon nanorod. We trap its centre-of-mass and align it along the linear polarization of the laser field. The rod can be set into rotation at a predefined frequency by exploiting the radiation pressure exerted by elliptically polarized light. The rotational motion of the rod dynamically modifies the optical potential, which allows tuning of the rotational frequency over hundreds of Kilohertz. Through nanofabrication, we can tailor all of the trapping frequencies and the optical torque, achieving reproducible dynamics which are stable over months, and analytically predict the motion with great accuracy. This first demonstration of full ro-translational control of nanoparticles in vacuum opens up the fields of rotational optomechanics, rotational ground state cooling and the study of rotational thermodynamics in the underdamped regime.

\end{abstract}

\maketitle

{\it Introduction---} Nanofabrication has advanced all areas of science, technology and medicine \cite{Bhushan2010}, including the field of optomechanics, where the motion of a mechanical oscillator is controlled by light. The quantum ground state of motion has been reached in optomechanical crystal devices \cite{Chan2011}, and superconducting microwave circuits \cite{Teufel2011}. Ground-state cooling enables the coherent transduction of signals \cite{Bochmann2013}, the production of non-classical states of light and matter \cite{Riedinger2016}, and the ultra-sensitive detection of motion \cite{Arcizet2006} and forces \cite{Krause2012}. Coherent optomechanical technology is limited by the coupling between the mechanical device and its environment, which leads to decoherence of quantum states, and by a reduction in mechanical quality factor due to clamping forces on the oscillator.

These limitations can be overcome by optically levitating the mechanical system, such that it oscillates in a harmonic trapping potential. Optical trapping is applicable from atoms in vacuum \cite{Phillips1998,Schlosser2001}, to complex organisms in liquid \cite{Ashkin1987}. By optically levitating nanoscale objects in vacuum, ultra-high mechanical quality factors ($Q \sim 10^{12}$) are predicted \cite{Chang2010}, and it may be possible to generate macroscopic quantum superpositions \cite{Romero-Isart2010}. Such massive quantum systems could test the limits of quantum physics \cite{Arndt2014,Bateman2014}, looking for the existence of new mechanisms of wave-function collapse such as spontaneous localisation \cite{Ghirardi1990} or gravitational effects \cite{Diosi1987,Ghirardi1990a,Penrose1996}. These goals require a high degree of control over all of the dynamics of the nanoparticle. 

%%%%%%%%%%%%%%%%%%
\begin{figure}[!ht]
	 {\includegraphics[width=0.46\textwidth]{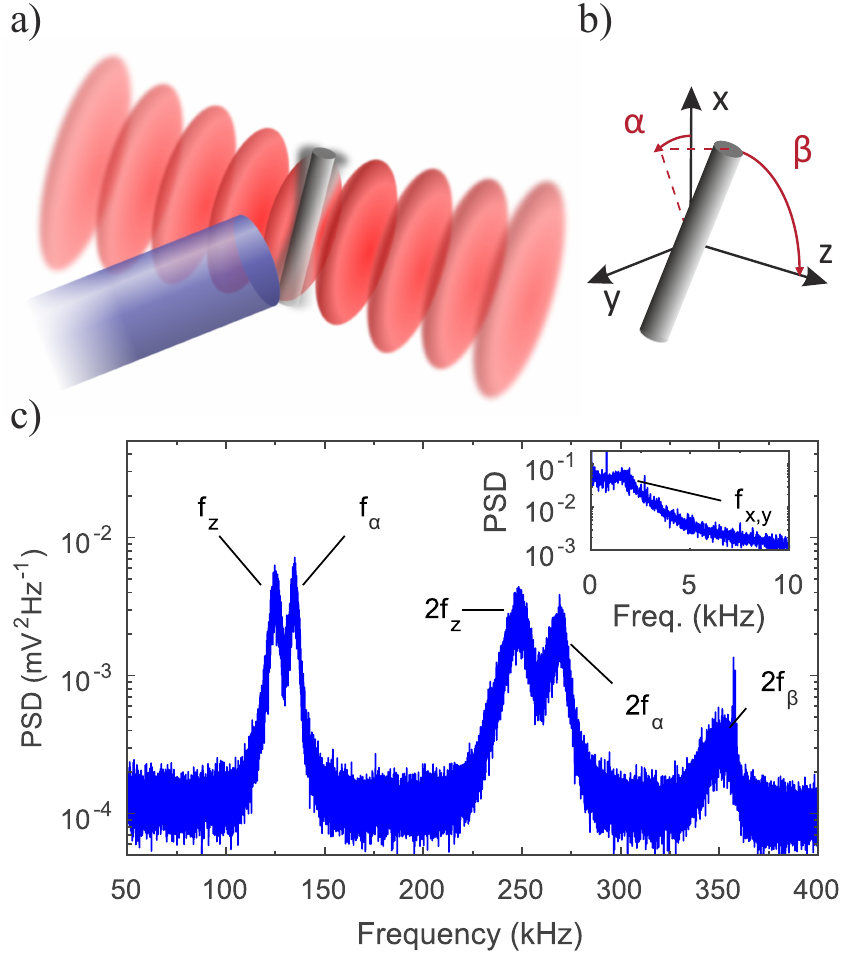}}	
\caption{\label{fig:exp} 
a) Nanofabricated silicon nanorods of length $\ell \simeq$(725 $\pm$15)\,nm and diameter $d \simeq$ (130 $\pm$13)\,nm are optically levitated in a standing laser wave at low pressures. The light they scatter is collected by a multimode optical fiber placed close to the trap waist. b) The rods have five degrees of freedom which can be controlled; three translational ($x,y,z$) and two rotational ($\alpha,\beta$). c) By monitoring the scattered light, trapping of all five degrees of freedom can be observed in the power spectral density (PSD) when the trap light is linearly polarized. This data was acquired at a pressure of 4\,mbar. The appearance of the various harmonics can be explained by slight misalignment of the trap as discussed in Supplementary Information 1.}
\end{figure}
%%%%%%%%%%%%%%%%%%% 

The field of levitated optomechanics is growing rapidly, with progress including feedback \cite{Li2011,Gieseler2012,Vovrosh2016} and cavity cooling \cite{Kiesel2013,Asenbaum2013,Millen2015,Fonseca2015} to the milli-Kelvin level and below, the sensing of forces on the zepto-Newton scale \cite{Ranjit2016}, and the study of Brownian motion \cite{Li2010} and equilibration \cite{Gieseler2014} in the underdamped regime. Experiments are often limited by the quality of commercially available nanoparticles. Impurities lead to absorption of the trapping light, causing loss at low pressures \cite{Millen2014}, and even graphitization of levitated diamond \cite{Rahman2015}. Recently, rotation has been detected in levitated particles \cite{Kane2010,Arita2013,Kuhn2015,Hoang2016}, displaying far greater rotation rates than experiments in liquid \cite{Friese1998, Tong2010, Lehmuskero2013, Shao2015}.

In this work, we trap clean, nanofabricated silicon nanorods, and study their center-of-mass and rotational motion. Our particles are of uniform, tailored size and shape, allowing a high degree of repeatability, predictability and control of the dynamics. We are able to trap the nanorods, trap and control their orientation, and tunably spin them using the radiation pressure exerted by the light field. While rotational control has been achieved in liquid \cite{Friese1998, Borghese2008, Jones2009, Tong2010, Lehmuskero2013}, this is the first demonstration of trapping nanorods in vacuum, and we observe novel features such as shape enhanced light-matter interactions and dynamic reshaping of the trapping potential. Such full control opens the way to optomechanical rotational cooling \cite{Bhattacharya2007, Stickler2016, Hoang2016}, even to the quantum level \cite{Stickler2016}.

\section{Experimental setup}

A single silicon nanorod is optically trapped in the focus formed by two counterpropagating laser beams of wavelength $\lambda~=~1550$\,nm, see Fig.~\ref{fig:exp}a). At this wavelength, silicon exhibits a high relative permittivity, $\varepsilon_{\rm r} = 12$, and negligible absorption, which is supported by the fact that we see no signature of heating due to light absorption (following the method in Ref. \cite{Millen2014}). The nanorods are tailored to have a length of $\ell = (725 \pm 15)$\,nm and a diameter of $d = (130 \pm 13)$\,nm, corresponding to a mass $M = (1.3 \pm 0.3) \times 10^{10}$\,amu. They are fabricated onto a silicon chip following the methods described in Ref. \cite{Kuhn2015}. The laser trap is characterized by a beam waist radius $w_0 \approx 27\,\mu$m and the total power $P_{\mathrm{tot}} = 1.35$\,W, making a large volume trap to enhance the rate of capture. The nanorods are trapped in a clean N$_2$ environment at a pressure of $p_{\rm g} = 4$\,mbar, after being launched by laser desorption from a silicon wafer, see Refs. \cite{Kuhn2015, SPIE2016}. Up to 10 nanorods are simultaneously trapped, and we perturb the trapping field until a single nanorod remains. The rods can be stably trapped for months at any pressure above 1\,mbar, below which they are lost, as observed in experiments with spherical nanoparticles \cite{Kiesel2013,Millen2014}.

%%%%%%%%%%%%%%%%%%
\begin{figure}[t]
	 {\includegraphics[width=0.46\textwidth]{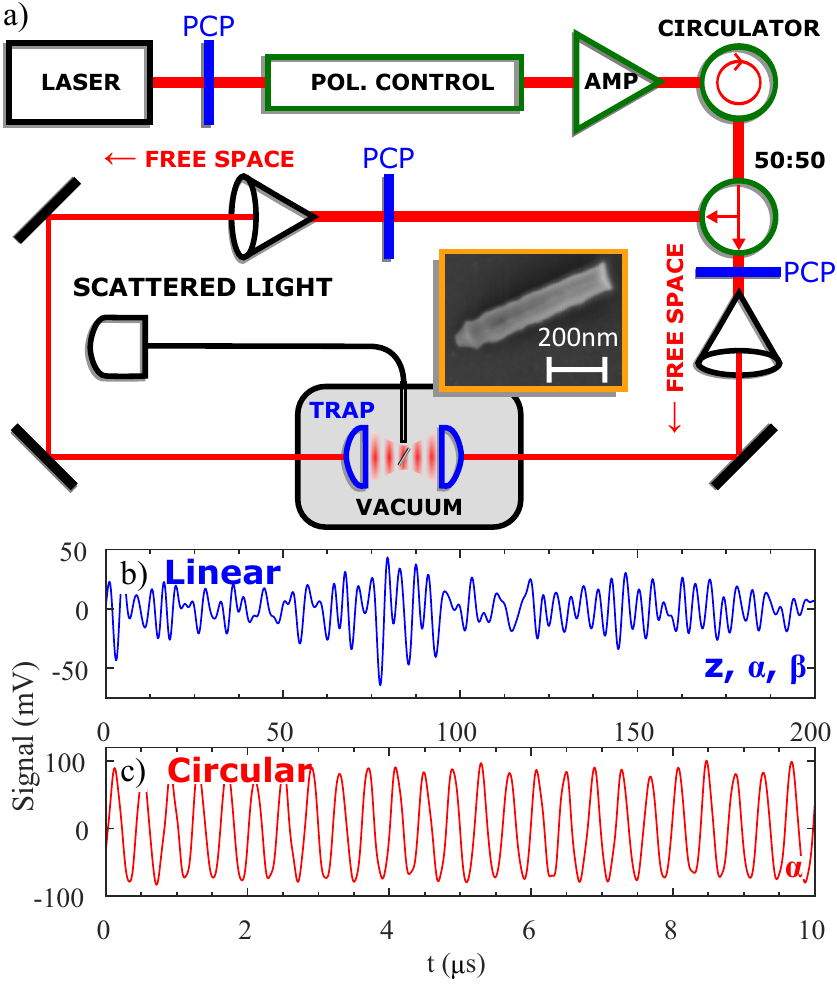}}	
\caption{\label{fig:layout} 
a) Experimental setup. Light at $\lambda = 1550$\,nm is produced by a fiber laser (Keysight 81663A), and then goes through an electro-optical in-fiber polarization controller (EOSPACE), allowing us to realise arbitrary waveplate operations. The light is amplified in a fiber amplifier (Hangzhou Huatai Optic HA5435B-1) and split equally to make the two arms of the trap. Stress induced birefringence in the fibers can be accounted for with polarization controlling paddles (PCP). The system is completely fiber-based until out-coupled to the aspheric trapping lenses (f=20\,mm). The inset shows an SEM micrograph of a rod that was launched and captured on a sample plate. The scattered light signal reveals the nanorod dynamics in case of b) co-linear polarization, and c) the strongly driven rotation of the rod for circularly polarized trapping light. 
}
\end{figure}
%%%%%%%%%%%%%%%%%%%

\section{Trapping the nanorods}

The motional state of the nanorod is described by its center-of-mass position $(x,y,z)$ and by its orientation $(\alpha,\beta)$, see Fig.~\ref{fig:exp}b), where $x$ points counter-parallel to the direction of gravity and $z$ along the beam axis. The orientation of the rod is parametrized by $\alpha$, the angle between the $x$-axis and the projection of the rod onto the $x$-$y$-plane, and $\beta$  the angle between the rod's symmetry axis and the beam propagation axis. The motion of the nanorod is measured via the light that the rod scatters out of the trap, which is collected with a 1\,mm diameter multimode optical fiber as described in Ref. \cite{Kuhn2015}.

The polarization of the two trapping beams determines the properties of the optical trap. In the case of co-linear polarization the rod aligns with the field polarization and is thus trapped in all its degrees of freedom. The resulting trapping frequencies can be measured in the power spectral density (PSD) of the scattered light signal, as shown in Fig.~\ref{fig:exp}c). Using a LiNb-polarization controller, we can perform arbitrary wave-plate operations on the polarization of the trapping light \cite{VanHaasteren1993}. The optical setup (see Fig. \ref{fig:layout}a)\,) is designed such that the rod experiences the same polarization from both arms of the counterpropagating trap. By realising a half-waveplate operation on the linearly polarized trapping beam we can align the rod along any direction orthogonal to the trap axis, as has been observed in liquid~\cite{Tong2010}.

The trapping frequencies of a harmonically captured rod can be calculated as \cite{Stickler2016}
\begin{eqnarray}\label{eqn:trapped}
f_{x,y} &= &\frac{1}{2\pi }\sqrt{\frac{8 P_{\rm tot} \chi_{\|}}{ \pi \varrho c w_0^4}}, \quad f_z = \frac{1}{2\pi} \sqrt{\frac{4 P_{\rm tot} \chi_{\|} k^2}{\pi \varrho c w_0^2 }}, \notag \\
f_{\beta} & = & \frac{1}{2\pi} \sqrt{\frac{48 P_{\rm tot} \chi_{\|}}{ \pi \varrho c w_0^2 \ell^2} \left( \frac{\Delta\chi}{\chi_{\|}} + \frac{(k \ell)^2}{12}\right)}, \notag \\
f_{\alpha} &= &\frac{1}{2\pi } \sqrt{\frac{48 P_{\rm tot} \Delta\chi}{ \pi \varrho c w_0^2 \ell^2}},
\end{eqnarray} 
where $k = 2 \pi / \lambda$, $\varrho = 2330$\,kg\,m$^{-3}$ is the density of silicon, $\chi_{\|} = \varepsilon_{\rm r} - 1$ is the susceptibility along the rod's symmetry axis and $\Delta\chi = (\varepsilon_{\rm r} - 1)^2 / (\varepsilon_{\rm r} + 1)$ is the susceptibility anisotropy \cite{Dehulst}. At the maximum input power we measure $f_{x,y} = (1.6 \pm 0.2)$\,kHz, $f_{z} = (124 \pm 1)$\,kHz, $f_{\alpha} = (134 \pm 1)$\,kHz and $f_{\beta} = (175.0 \pm 0.5)$\,kHz, see Fig.~\ref{fig:exp}c). For comparison, a silicon nanosphere of the same volume under the same experimental conditions would have $f_{z} = 58$\,kHz, and a silica sphere would have $f_{z} = 47$\,kHz, illustrating the great potential for silicon nanorods in cavity cooling experiments \cite{Stickler2016}. We can use the measured frequencies to deduce the trapping waist radius $w_0 = (27 \pm 3)\,\mu$m, which is the only free experimental parameter. The measured frequencies agree well with the theoretical expectations, as shown in Fig.~\ref{fig:circ}c). The slight ($<5\%$) discrepancy between the measured and predicted value of $f_{\beta}$ is attributed to the fact that the rods have finite diameter and the generalized Rayleigh-Gans approximation~\cite{Stickler2016} is not strictly valid.

\section{Spinning the nanorods}

When the trapping light is circularly polarized, the trapping potential for the $\alpha$ motion vanishes whilst the standing wave structure along $z$ is retained. The radiation pressure of the laser field exerts a constant torque $N_\alpha$ acting on $\alpha$. Adapting the theory presented in Ref. \cite{Stickler2016}, the resulting torque is obtained as
\begin{align}
\label{eq:torque}
N_\alpha = \frac{P_{\rm tot}\Delta\chi \ell^2 d^4 k^3}{48 c  w_0^2}  \left [\Delta \chi \eta_1(k \ell) +\chi_\bot \eta_2(k \ell) \right ],
\end{align}  
where the two functions $\eta_{1,2}(k \ell)$ are given by

\begin{eqnarray}
\eta_1(k \ell) & = & \frac{3}{4} \int_{-1}^1 \mathrm{d}\xi~(1 - \xi^2) \mathrm{sinc}^2 \left ( \frac{k \ell \xi}{2} \right ), \notag\\
\eta_2(k \ell) & = & \frac{3}{8} \int_{-1}^1 \mathrm{d}\xi~(1 - 3\xi^2) \mathrm{sinc}^2 \left ( \frac{k \ell \xi}{2} \right ).
\end{eqnarray}
For short rods, $k \ell \ll 1$, one has $\eta_1 \simeq 1$ while $\eta_2 \simeq 0$.

Collisions with residual gas molecules lead to damping of the rotational motion. Since the mean free path of the gas molecules exceeds the diameter of the rod (free molecular regime \cite{Cercignani1975}), the rotational damping rate for diffuse reflection of gas molecules of mass $m_{\rm g}$ takes the form \cite{Eisner1981}
\begin{align} \label{eq:drate}
\Gamma =  \frac{d \ell p_{\rm g}}{2 M} \sqrt{\frac{2 \pi m_{\rm g}}{k_{\rm B} T}} \left(\frac{3}{2} + \frac{\pi}{4}\right),
\end{align} 
where $T$ is the gas temperature.

%%%%%%%%%%%%%%%%%%
\begin{figure*}[ht]
	 {\includegraphics[width=0.98\textwidth]{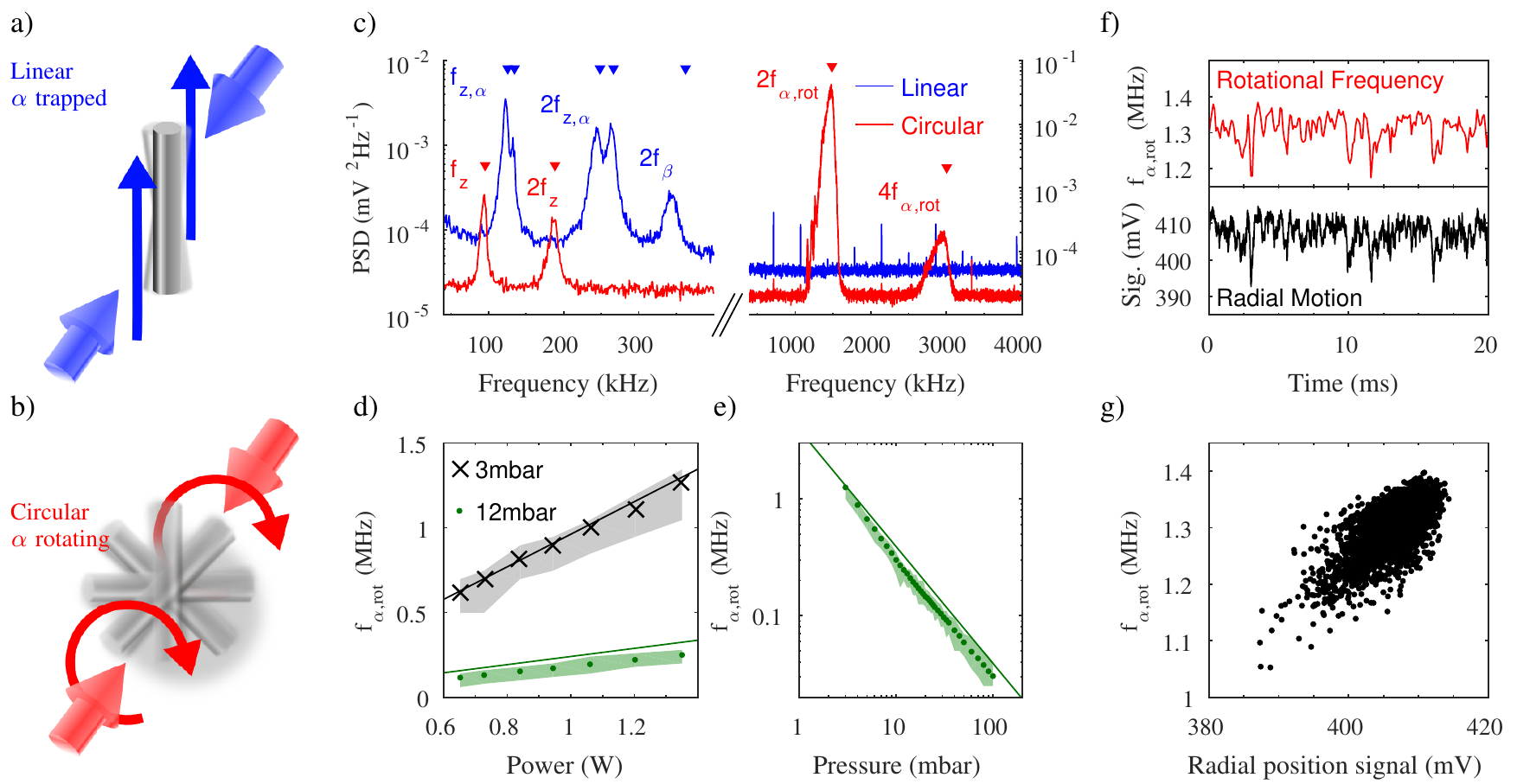}}	
\caption{\label{fig:circ} 
Comparing the dynamics when the nanorod is a) trapped in all degrees of freedom by linearly polarized light and b) driven to rotate in the $\alpha$ direction by circularly polarized light. c) The PSD for circularly (red) and linearly (blue) polarized light. For circular polarization, the trapped frequency $f_{\alpha}$ vanishes, and the rotational frequency $f_{\alpha,\mathrm{rot}}$ appears. The peak at $f_{\beta}$ vanishes since the motion in $\beta$ is stabilized when the rod is spinning. Markers indicate predicted trapping frequencies. The rotational frequency scales d) linearly with power, and e) decreases with increasing pressure, as predicted by Eq.~(\ref{eqn:rot}). Markers represent the mean value of $f_{\alpha,\mathrm{rot}}$, the shaded areas represent the full range of $f_{\alpha,\mathrm{rot}}$, and solid lines are the theoretically expected maximal value of $f_{\alpha,\mathrm{rot}}$. The broad frequency distribution of $f_{\alpha,\mathrm{rot}}$ is due to coupling between the motion in $\alpha$ and $x,y$ (radial). f) Perturbations from the equilibrium position (lower panel) are reflected in instantaneous frequency fluctuations (top panel). g) The correlation between the radial position and $f_{\alpha,\mathrm{rot}}$.}
\end{figure*}
%%%%%%%%%%%%%%%%%%%

The maximum steady-state rotation frequency is obtained by balancing the torque Eq.~\ref{eq:torque} with the damping Eq.~\ref{eq:drate},
\begin{align}
\label{eqn:rot}
f_{\alpha,{\rm max}} = \frac{N_\alpha}{2 \pi I \Gamma},
\end{align}
with $I = M \ell^2 /12$ the rod's moment of inertia. This expression agrees well with the measured value of the rotation frequency $f_{\alpha,\mathrm{rot}}$ as a function of power and pressure, as shown in Figs.~\ref{fig:circ}d) and e), respectively.

A comparison of the PSD for the co-linear and the circular polarization traps is shown in Fig.~\ref{fig:circ}c). The peak related to the trapping frequency at $f_\alpha$ vanishes and a pronounced peak at $2 f_{\alpha, {\rm rot}}$ arises. We are only sensitive to $2 f_{\alpha, {\rm rot}}$ due to the symmetry of the rod. The rotation of the rod in the circularly polarized field results in a reduced average susceptibility and thus a weaker trapping potential, which shifts the axial trapping frequency to $f_{z,{\rm rot}} = 94$\,kHz as discussed in Supplementary Information 1. The rapid rotation in $\alpha$ leads to a stabilization in $\beta$ and hence the complete suppression of the peak at $f_\beta$ in Fig.~\ref{fig:circ}c). A similar effect has also been observed for spinning microspheres \cite{Arita2013}. 

The broad distribution of frequencies about $2 f_{\alpha, {\rm rot}}$ is due to perturbations temporarily decreasing the rotation rate, which then takes time to spin back up to the maximum value. For example, irregular excursions in the radial $x,y$ directions lead to variations in the instantaneous rotation frequency via variation in the local light intensity, as shown in Fig.~\ref{fig:circ}f), with the correlation clearly shown in Fig.~\ref{fig:circ}g). The maximum rotation rate is limited by pressure in this set-up, with an ultimate limit presumably set by material properties. In previous work, rotation rates of 50\,MHz were observed for free nanorods in UHV \cite{Kuhn2015}.

%%%%%%%%%%%%%%%%%%
\begin{figure}[t]
	 {\includegraphics[width=0.46\textwidth]{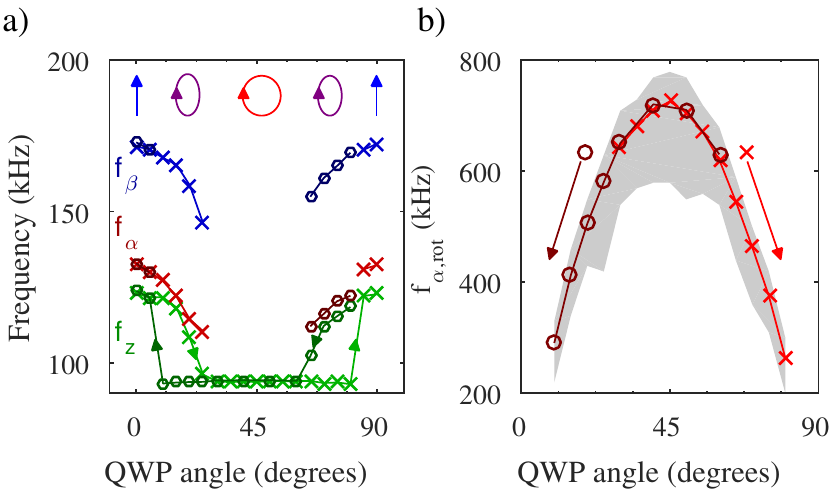}}	
\caption{\label{fig:QWP} 
Effect of performing a quarter-waveplate (QWP) operation on the trapping light at 5\,mbar, either increasing from from $0^\circ$ (crosses) or decreasing from $90^\circ$ (circles). At $0^\circ$ and $90^\circ$ the trap is linearly polarized along the $y$-axis. At $45^\circ$ the polarization is circular. a) Shift of the trapping frequencies for different QWP settings.  For small deviations from linear polarization the trapping frequencies decrease due to a lower trapping potential. At $30^\circ$ from the starting linear polarization, the light is circularly polarized enough to drive $f_{\alpha, {\rm rot}}$, at which point $f_{\alpha,\beta}$ vanish, and $f_z$ drops. At $85^\circ$ from the starting linear polarization, the motion becomes trapped again. b) Due to this hysteresis the driven rotational frequency $f_{\alpha, {\rm rot}}$ can be tuned over several hundred Kilohertz via the ellipticity of the trapping field. The markers indicate the mean value of $f_{\alpha,\mathrm{rot}}$, and the shaded region represents the range of measured frequencies.}
\end{figure}
%%%%%%%%%%%%%%%%%%%

\section{Tuning the rotational frequency}

To study the effect of driven rotation in more detail we use the polarization controller to perform a quarter-waveplate (QWP) operation on the trapping light and track the motion of the rod at each setting, see Fig.~\ref{fig:QWP}. Starting from a linear polarization along $x$ and increasing its ellipticity at first leads to a shift of all trapping frequencies to lower values due to a reduced trap depth, as shown in Fig.~\ref{fig:QWP}a). At a QWP setting of $30^\circ$ the radiation pressure induced torque starts driving the rod into rotation over the trapping potential in the direction of $\alpha$, the frequencies $f_{\alpha,\beta}$ vanish, and $f_z$ drops to a steady value of $f_{z, {\rm rot}} = (94 \pm 1)$\,kHz, as also seen in Fig.~\ref{fig:circ}c). Rotating beyond $45^\circ$, one may expect the nanorod to become trapped again at $60^\circ$, however the rod is not trapped until $85^\circ$. When starting at $90^\circ$ and decreasing the QWP angle, the rod spins at $60^\circ$, and is not trapped until $5^\circ$, showing a symmetric hysteresis, see Fig~\ref{fig:QWP}b).

This effect is due to the anisotropy of the susceptibility tensor: A trapped rod experiences the full trap depth related to $\chi_\|$ whereas the trapping potential for a spinning rod is proportional to the susceptibility averaged over rotations in the 2D plane orthogonal to the beam axis $(\chi_\|+\chi_\bot)/2$, which is smaller by a factor of 1.7. Thus, it requires a greater torque to spin a trapped rod than to maintain the rotation of an already spinning rod. The value of $f_{\alpha, {\rm rot}}$ varies with the ellipticity of the light, as shown in Fig.~\ref{fig:QWP}b). By exploiting the dynamical modification of the trap depth we can extend the range over which the rotation frequency can be tuned to many hundreds of Kilohertz.

\section{Conclusions}

In summary, we present a method to capture and levitate nanofabricated silicon nanorods at low pressures, working with telecoms wavelengths in a fibre-based setup. We can precisely control the length and diameter of our nanorods, meaning we can tailor rods to attain particular trapping and rotational frequencies. We are able to trap all relevant degrees of freedom, and control the orientation of the rods via the polarization of the trapping beams. By using circularly polarized light we can spin the nanorods at more than 1\,MHz, and tune this frequency over hundreds of Kilohertz by introducing ellipticity into the field polarization and through a dynamic modification of the trapping potential. When the rod is spinning we notice a stabilization of the tilt angle $\beta$ and a coupling to the radial motion $x,y$. The system is very well understood as shown by the excellent agreement between experiment and theory. The high degree of control opens the way to study rotational optomechanics \cite{Rubin2011,Bhattacharya2015,Shi2015}, orientational decoherence \cite{Stickler2016b, Zhong2016}, rotational underdamped Brownian motion and stochastic thermodynamics, and synchronisation of multiple rotors due to optical binding \cite{Simpson2016}. This is the first use of silicon in an optical trap in vacuum, and its high susceptibility and low absorption in this frequency band, combined with the shape-enhanced susceptibility of rods, will enable rotational cavity cooling to the quantum level \cite{Stickler2016,Hoang2016}. Such deeply trapped, cooled particles may be used as point sources for orientation dependent interference experiments \cite{Shore2015,Stickler2015}.

\section*{Funding Information}
We are grateful for financial support by the Austrian Science Funds (FWF) in the project P27297 and DK-CoQuS (W1210-3).  J.M. acknowledges funding from the European Union’s Horizon 2020 research and innovation programme under the Marie Sk{\l}odowska-Curie grant agreement No 654532. F.P. acknowledges the Legacy Program (Israel Science Foundation) for its support.

\section*{Acknowledgments}
We acknowledge support by S. Puchegger and the faculty center for nanostructure research at the University of Vienna in imaging the nanorods. We thank Frank for his dedication to the project.

\bigskip \noindent See \href{link}{Supplement 1} for supporting content.

% Bibliography

\pagebreak
\onecolumngrid

\begin{center}
\textsc{\LARGE Supplementary Information}\\[1.2cm]
\end{center}

\setcounter{equation}{0}
\setcounter{figure}{0}
\setcounter{table}{0}
\setcounter{page}{1}
\makeatletter
\renewcommand{\theequation}{S\arabic{equation}}
\renewcommand{\thefigure}{S\arabic{figure}}
\renewcommand{\bibnumfmt}[1]{[S#1]}
\renewcommand{\citenumfont}[1]{S#1}

\section{Distinguishing different translational and rotational motions}
\label{sec:alpha}

\begin{figure}[b]
   \begin{center}
      \includegraphics{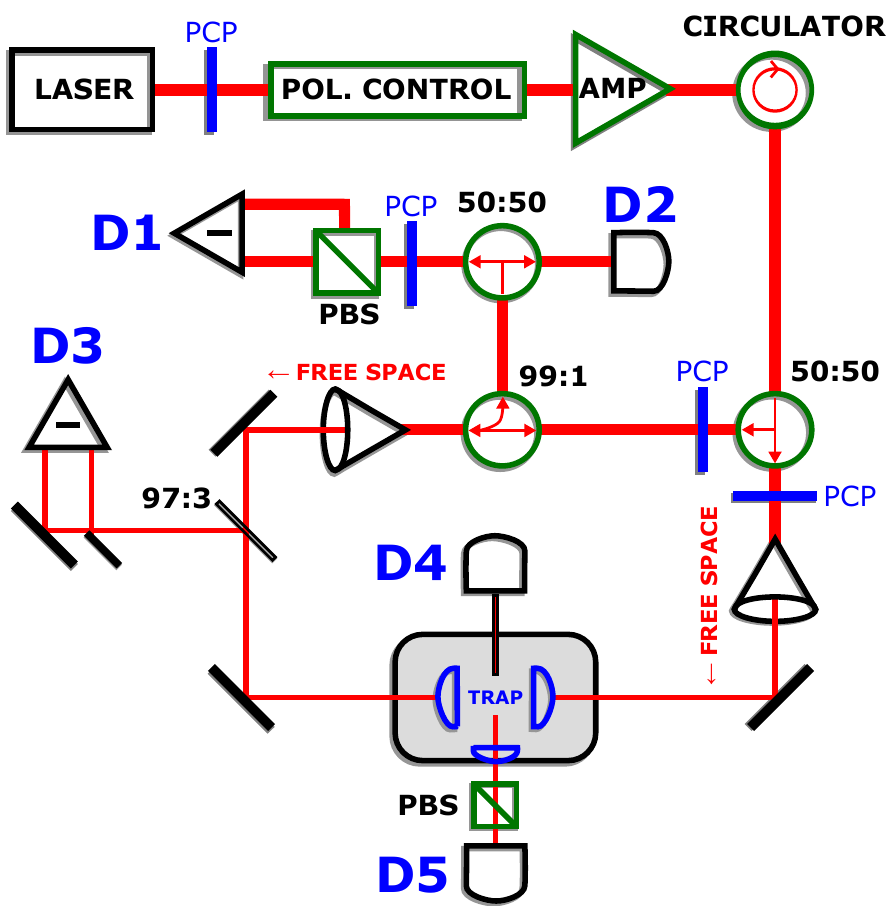}
      \end{center}
   \caption[schem] 
   { \label{fig:layout} 
The full optical layout, including mechanisms for independently measuring different trapping frequencies. A small portion of the free-space light is split on a D-shaped mirror, yielding the radial motion $x$ on detector D3. A portion of the light that is coupled back into the fibers is picked off, and split equally to be sent to two detectors. D2 monitors the motion in $z$, as described in Ref.~\cite{Gieseler12}, and D1 the motion in $\alpha$. We collect the light which the nanorod scatters using a multimode fiber, and monitor its intensity on detector D4, and use a lens to collect the scattered light to perform a polarization sensitive measurement in the 45$^{\circ}$ basis on detector D5.}
  \end{figure}

All trapping frequencies of the nanorods can be calculated, based on their geometry as measured through scanning electron microscopy. The scattered light signal contains information about all degrees of freedom, with a high signal-to-noise ratio (SNR)~\cite{Kuhn2015}. In the manuscript, all data for $(x,y,z,\alpha)$ is extracted from the scattered light detector D4, and the motion in $\beta$ is extracted from a polarization dependent measurement on the scattered light D5. 

We independently check all motional degrees of freedom to confirm our assignation of frequencies, using other detectors. This information is not used in the manuscript. The full optical layout is shown in Fig.~\ref{fig:layout}, and each detector and its uses are outlined in table~\ref{tab:detectors}.

\begin{figure*}[b]
   \begin{center}
      \includegraphics{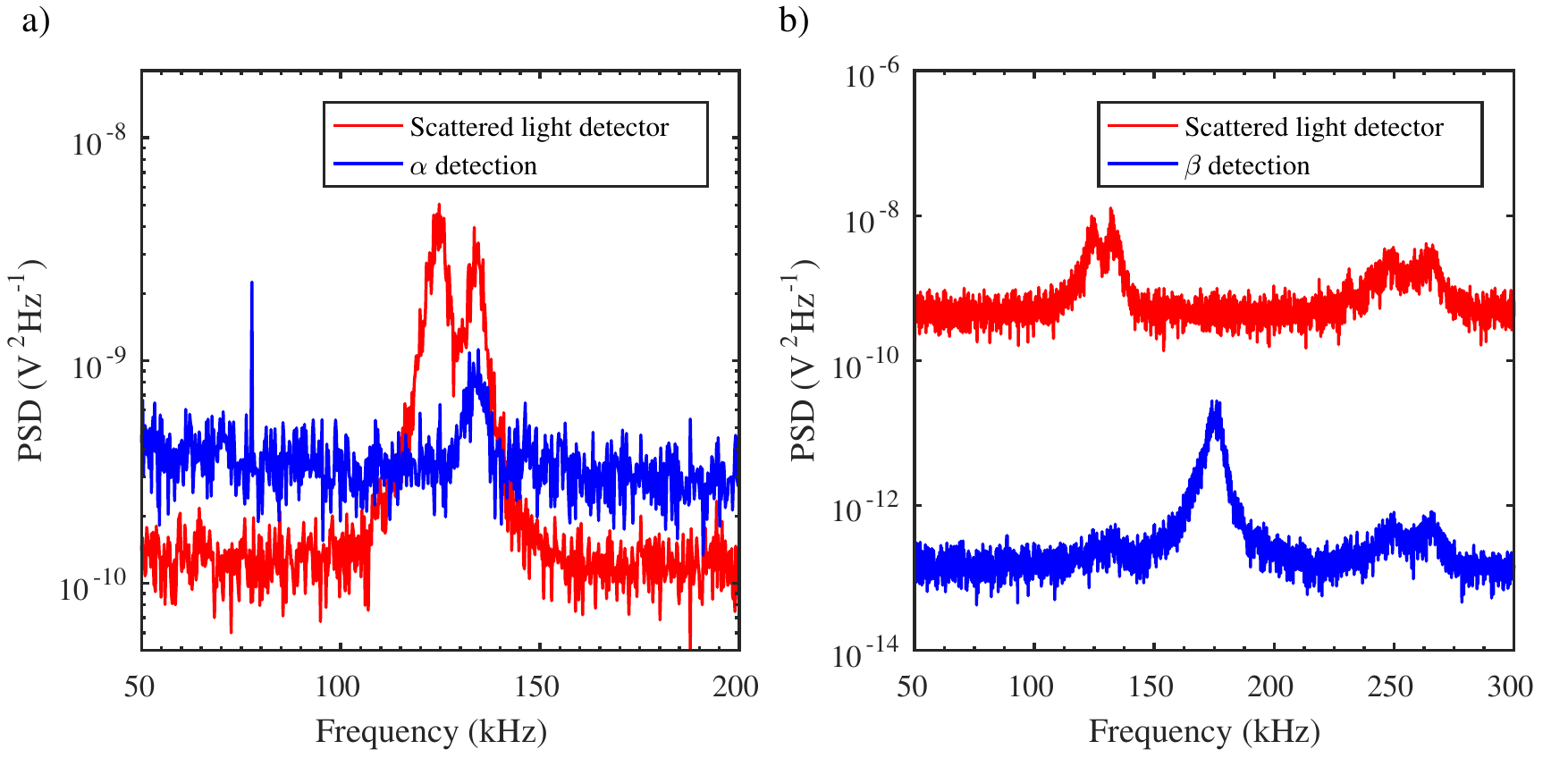}
      \end{center}
   \caption[schem] 
   { \label{fig:alpha} 
a) A comparison of the PSD of the signal from the scattered light detector D4 (red), which is sensitive to motion in all directions, and the polarization sensitive detector D1 (blue), which is only sensitive to $\alpha$. This confirms that the higher frequency peak corresponds to the $\alpha$ motion, which agrees with our calculations. b) A comparison of the PSD from the scattered light detectors D4 (red) and D5 (blue, polarization sensitive). D5 is sensitive to the frequency $f_\beta$, which isn't visible on detector D4.}
  \end{figure*}

\textbf{Motion in $z$ and $\alpha$:} We expect the trapping frequencies for $z$ and $\alpha$ to be $f_z = 124$\,kHz and $f_{\alpha} = 134\,$kHz respectively. In the data we see a double-peak in the Power Spectral Density (PSD) of the scattered light signal D4 around these frequencies, as shown in Fig.~\ref{fig:alpha}a). To confirm experimentally which peak is which we implement an additional detection scheme. The nanorod rotates the polarization of the trapping light, depending on the angle it makes to the polarization axis (i.e. by an amount proportional to $\alpha$).

We monitor the trapping light that has interacted with the nanorod by collecting some of the light coupled back into the fiber outcouplers, splitting it off with a 99:1 fiber beamsplitter and turning its linear polarization by 45$^{\circ}$. This light then goes through a fiber polarizing beamsplitter (PBS), and each arm of the PBS is coupled onto a fast fiber-coupled balanced detector D1 (Thorlabs PDB420C). By this method we measure just the rotation in $\alpha$, as shown in Fig.~\ref{fig:alpha}a). This confirms that the higher frequency peak is due to motion in $\alpha$.

\begin{table*}[tp]
	\centering
		\begin{tabular}{l | p{10cm} | c}
			Detector & Model & Motional sensitivity\\ \hline
			D1 & Fiber coupled differencing photodetector, Thorlabs PDB420C & $\alpha$ \\
			D2 & Fiber coupled photodetector Bookham PP-10GC58L & $x,y,z$ \\
			D3 & Homebuilt differencing photodetector & $x$ or $y$ \\
			D4 & Multimode fiber coupled Hamamatsu photodiode G12180-005A with Femto HCA-100M-50K-C amplifier & All degrees of freedom \\
			D5 & Free space photodetecotr Thorlabs DET10C & $\beta$ \\
		\end{tabular}
	\caption{Detectors used to measure the motion of a levitated silicon nanorod. Only detectors D4 and D5 are used in the manuscript, D1-3 are used for confirming our assignation of motional degrees of freedom.}
	\label{tab:detectors}
\end{table*}

We also monitor the intensity fluctuations of this picked-off light on detector D2, which yields information about the motion in $z$, as described in Ref.~\cite{Gieseler12}. The quality of this signal is poor, due to the large beam waist of our trap, and could be improved through a difference measurement. It reconfirms that the peak at 124\,kHz is due to motion in $z$. 

\textbf{Motion in $x,y$:} We pick off a small amount of the light that has interacted with the nanorod using a free-space 97:3 beamsplitter. This light is incident on a D-shaped mirror to cut the beam in half, and the resulting two beams are measured on a differencing photodiode D3. This yields the motion in $y$, as described in Ref.~\cite{Gieseler12}. By performing a half-waveplate operation on the trapping light we can rotate the nanorod to measure $x$. Due to the large beam waist of our trap, the SNR is poor for this signal.

\textbf{Motion in $\beta$:} Using a lens we collect the scattered light emitted in the opposite direction to that collected by the multimode fiber. This light is sent through a polarizing beamsplitter cube, which is rotated by 45$^{\circ}$ with respect to the $x$-axis, and monitored with a photodiode D5. This yields information about the motion in $\beta$, as shown in Fig.~\ref{fig:alpha}b).

\subsection*{Further notes on detection}

The SNR from the scattered light detector D4 is significantly better than that from detectors D1,2,3. Because of this, even though the $z$ and $\alpha$ peaks overlap, fitting the PSD of D4 is the most accurate method for monitoring the dynamics of the nanorod. However, we can always use the other detectors to confirm our findings. The SNR of the scattered light signal is so good because we only collect light that is scattered by the nanorod, with virtually zero background.

We expect the scattered light signal D4 to be only sensitive to the first harmonic of all motions $2f_{x,y,z,\alpha,\beta}$, since it depends on position squared. However, Fig.~\ref{fig:alpha}a) shows that we measure the fundamental motions $f_{z,\alpha}$ on D4. This is due to a slight misalignment between the trapping beams, and because the polarization in each arm of the trap is not perfectly identical. We confirm this through calculation, which also confirms that D4 is not sensitive to the fundamental frequency $f_\beta$. 

The polarization (rather than intensity) sensitive detectors D1,5 are sensitive to the sign of the motion in $\alpha,\beta$ respectively, and so the PSDs in Fig.~\ref{fig:alpha} show the frequency $f_{\alpha,\beta}$ and not its harmonic.

\section{Fitting data}
\label{sec:fitting}

To extract trapping frequencies we fit the PSD of the scattered light time series with the function
\begin{equation}
\label{eqn:PSD}
\mathrm{PSD}(\omega) =  C_d^2 \frac{2 k_B T}{M_d}\frac{\Gamma_d}{ (\omega_d^2 - \omega^2)^2 + \omega^2\Gamma_d^2},
\end{equation}
where $d$ labels the degrees of freedom (i.e. $x,y,z,\alpha,\beta$), $\Gamma_d$ is the (angular) momentum damping rate, $\omega_d$ is the trapping frequency, $M_d$ is the particle's mass or moment of inertia and $C_d$ is the calibration between our measured signal and absolute motional information. The derivation of this PSD is standard, e.g.~\cite{Gieseler12, Millen14}. When recording the time series of the particle's motion, the PSD can usually be calibrated to extract $C$ and convert from units of V$^2$Hz$^{-1}$ to m$^2$Hz$^{-1}$. However, our signal contains both translational ($x,y,z$) and rotational ($\alpha,\beta$) information, so such a global calibration isn't possible. As an indication of sensitivity, the peak sensitivity for motion in the $z$ direction from the PSD from detector D4 is $3\,\mu\mathrm{m}^2/\mathrm{Hz}^{-1}$.

\begin{figure*} [b]
   \begin{center}
      \includegraphics{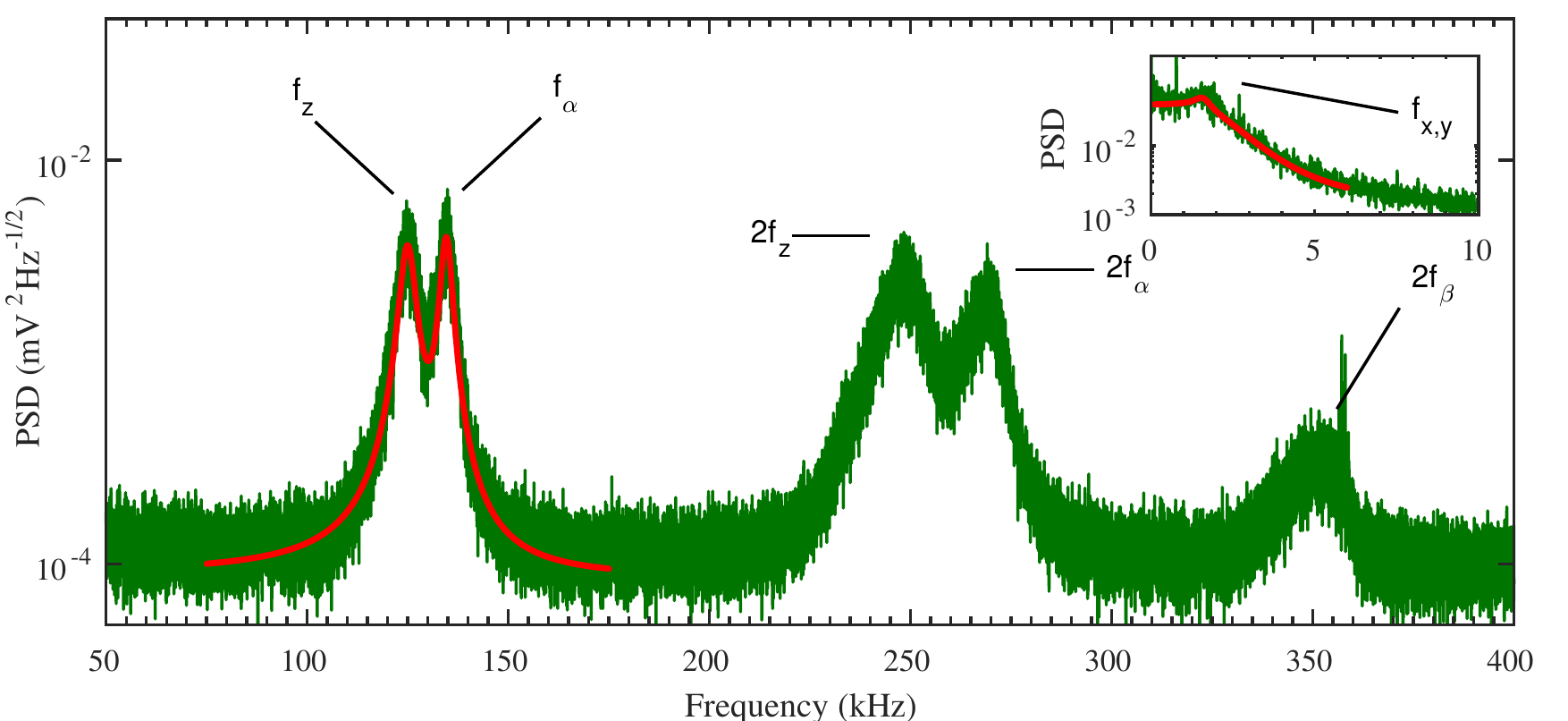}
      \end{center}
   \caption[schem] 
   { \label{fig:fitting} 
The PSD of the scattered light signal showing all motional degrees of freedom. Solid red lines are fits to the data using Eq.~\ref{eqn:PSD}.}
  \end{figure*} 

Each degree of freedom can be fit with this expression. Figure~\ref{fig:fitting} shows an example of fitting the PSD. Due to the proximity of the $z$ and $\alpha$ peaks in the scattered light PSD, we fit the data with a sum of two of the functions defined in Eq.~\ref{eqn:PSD} to extract the parameters for both the $z$ and $\alpha$ motions.

\section{Trapping frequency power scaling}

\begin{figure*} [ht]
   \begin{center}
      \includegraphics{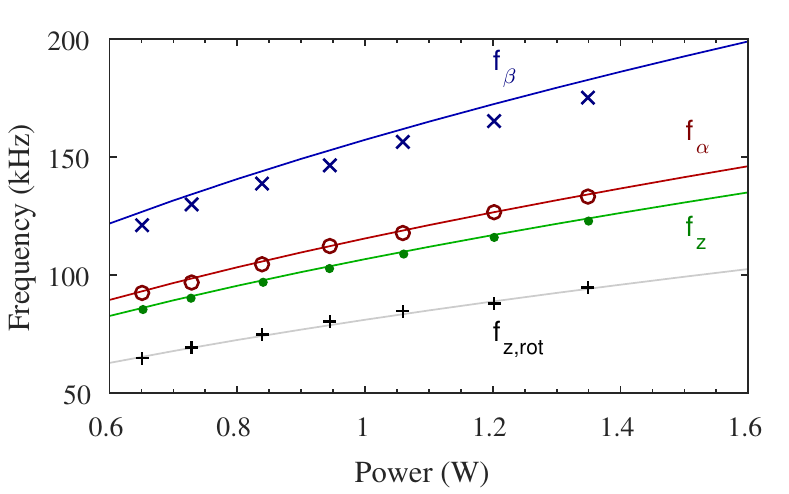}
      \end{center}
   \caption[schem] 
   { \label{fig:powpres} 
Variation in the trapping frequencies with total trap power, markers are data, solid lines are the theoretical predictions. The frequency $f_{z,\mathrm{rot}}$ is the trap frequency in the $z$ direction when the nanorod is rotating in the $\alpha$ direction. Experimental uncertainties are smaller than the data markers.}
  \end{figure*} 

We expect all trapping frequencies to scale with the square root of power, see Eq.~(1) of the manuscript. Figure~\ref{fig:powpres} shows the measured trapping frequencies as a function of power in comparison to the theoretical expectation. The only unknown experimental parameter is the laser waist $w_0$, which we extract from the ratio of $f_z$ to $f_x$. The ratio of $f_z$ to $f_{\alpha}$ confirms the length of our nanorod, which agrees with scanning electron microcope images.  We observe excellent agreement between theory and $f_{\alpha,z}$, and also for the axial frequency when the rod is rotating $f_{z,\mathrm{rot}}$. The discrepancy ($< 5~\%$) between theory and experiment for $f_\beta$ is attributed to the fact that the rods have finite diameter $(d  = 130~\mathrm{nm})$ and, the generalized Rayleigh-Gans approximation \cite{Stickler2016} is not strictly valid.

\section{Reduction of trapping potential when the particle is rotating}

Due to the rotation of the rod in the plane orthogonal to the trap axis, only the averaged susceptibility $(\chi_\| + \chi_\bot )/2$ enters the trapping potential of the translational motion in $z$-direction. This means that in Eq.~(1) of the manuscript, $\chi_\|$ has to be replaced by $(\chi_\| + \chi_\bot)/2$. This agrees well with the measured reduction of the trapping frequency, see Fig. \ref{fig:powpres}.

\section{Extracting information from the rotational motion}

When the trapping light is circularly polarized the nanorod rotates in the plane orthogonal to the trap axis ($\alpha$ direction) with the frequency $f_{\alpha,\mathrm{rot}}$. As shown in the paper, $f_{\alpha,\mathrm{rot}}$ has a broad distribution. To analyze this motion we extract the instantaneous frequency, using time bins 100 times longer than the mean rotational period. In the paper we present the mean value of $f_{\alpha,\mathrm{rot}}$ and display shaded regions around the data points representing the minimum and maximum values of $f_{\alpha,\mathrm{rot}}$. The theoretical analysis predicts the maximum rotation frequency $f_{\alpha,\mathrm{max}}$, which is determined by the balance between the torque exerted by the light field and the rotational friction due to collisions with gas molecules.

\end{document}